\def\Y{\Upsilon}
\def\B{{\cal B}}
\def\bb{b\bar b}
\def\chib{\chi_b}
\def\eonem#1#2{\ifmmode
\left| <\!#1|r|#2\!> \right|
\else%
$\left| <\!#1|r|#2\!> \right|$
\fi}
\def\etal{{\it et~al.}}
\def\ee{e^+e^-}
\def\mm{\mu^+\mu^-}
\def\LL{l^+l^-}

\long\def\simplex#1#2#3#4{
\begin{figure}[#1]
   \begin{center}
   \quad\\[-1cm]
   \quad
   \hbox{
   \quad 
   \parbox[t]{\hsize}{ \psfig{figure=#2,width=10.0cm} 
   \caption[]{\small  \label{fig:#3} #4 }}
   }
   \quad
   \end{center} 
\end{figure}
}
\long\def\ssimplex#1#2#3#4#5{
\begin{figure}[#1]
   \begin{center}
   \quad\\[-1cm]
   \quad
   \hbox{
   \quad 
   \parbox[t]{\hsize}{ \psfig{figure=#2,width=#5} 
   \caption[]{\small  \label{fig:#3} #4 }}
   }
   \quad
   \end{center} 
\end{figure}
}
\long\def\duplex#1#2#3#4#5#6#7{
\begin{figure}[#1]
   \begin{center}
   \quad\\[-1cm]
   \quad
   \hbox{
   \quad 
   \parbox[t]{7.5cm}{ \psfig{figure=#2,width=8.0cm} 
   \caption[]{\small  \label{fig:#3} #4 }}
   \quad
   \parbox[t]{7.5cm}{ \psfig{figure=#5,width=8.0cm} 
   \caption[]{\small \label{fig:#6} #7 }}
   }
   \quad
   \end{center} 
\end{figure}
}
\long\def\vduplex#1#2#3#4#5#6#7{
\begin{figure}[#1]
   \begin{center}
   \quad\\[-3.5cm]
   \quad
   \parbox[t]{\hsize}{ \psfig{figure=#2,width=10.0cm} 
   \caption[]{\small  \label{fig:#3} #4 }}
   \quad
   \\[-1cm]
   \quad
   \parbox[t]{\hsize}{ \psfig{figure=#5,width=10.0cm} 
   \caption[]{\small \label{fig:#6} #7 }}
   \quad
   \end{center} 
\end{figure}
}
\long\def\quadruplex#1#2#3#4#5#6#7#8#9{
\begin{figure}[#1]
   \begin{center}
   \quad\\[-1cm]
   \quad
   \hbox{
   \quad 
   \parbox[t]{7.5cm}{ \psfig{figure=#2,width=8.0cm} 
   \caption[]{\small  \label{fig:\1} #3 }}
   \quad
   \parbox[t]{7.5cm}{ \psfig{figure=#4,width=8.0cm} 
   \caption[]{\small \label{fig:\2} #5 }}
   }
   \\[-1cm]
   \hbox{
   \quad 
   \parbox[t]{7.5cm}{ \psfig{figure=#6,width=8.0cm} 
   \caption[]{\small  \label{fig:\3} #7 }}
   \quad
   \parbox[t]{7.5cm}{ \psfig{figure=#8,width=8.0cm} 
   \caption[]{\small \label{fig:\4} #9 }}
   }
   \quad
   \end{center} 
\end{figure}
}

\documentclass[aps,prd,preprint,superscriptaddress,tightenlines]{revtex4}
\usepackage{graphicx}
\usepackage{dcolumn}
\usepackage{bm}
\usepackage{epsfig}

\begin{document}

\preprint{CLEO CONF 02-07}
\preprint{ICHEP02 ABS949}

\title{Study of Two-Photon Transitions in CLEO-III $\Y(3S)$ Data}
\thanks{Submitted to the 31$^{\rm st}$ International Conference on High Energy
Physics, July 2002, Amsterdam}


\author{D.~Cinabro}
\author{M.~Dubrovin}
\author{S.~McGee}
\affiliation{Wayne State University, Detroit, Michigan 48202}                               
\author{A.~Bornheim}
\author{E.~Lipeles}
\author{S.~P.~Pappas}
\author{A.~Shapiro}
\author{W.~M.~Sun}
\author{A.~J.~Weinstein}
\affiliation{California Institute of Technology, Pasadena, California 91125}                
\author{R.~Mahapatra}
\affiliation{University of California, Santa Barbara, California 93106}                     
% ...
\collaboration{CLEO Collaboration}
\noaffiliation


\date{July 23, 2002}
\begin{abstract}
We have studied two-photon transitions from $\Y(3S)$ decays 
recorded by the CLEO-III detector
in exclusive events with two photons and either two electrons or two muons.
We obtain precision measurements of the $\chib(2P_J)$ masses for $J=2$ and
$J=1$.
The transition rates for all three spin states of the 
$2P$ triplet are measured with improved
precision which leads to a better determination of their hadronic width
ratios.
We also observe rare transitions via the $\chib(1P)$ states.
The measured rates for these transitions allow a determination of 
$<\!1P|r|3S\!>$, the E1 matrix element,
which is more sensitive to the structure of the $\bb$ states than
the $<\!2P|r|3S\!>$ matrix element 
which dominates the radiative decays of the $\Y(3S)$ state.

We also present first upper limits on the branching ratios for 
$\Y(3S)\to\pi^0\Y(2S)$, $\Y(3S)\to\pi^0\Y(1S)$ and
a new limit for the branching ratio for $\Y(3S)\to\eta\Y(1S)$. 
\end{abstract}

\maketitle

\section{Introduction}

Long-lived $b\bar b$ states are especially well suited for
testing QCD via lattice calculations \cite{LatticeQCD}
and effective theories
of strong interactions, like 
potential models \cite{PotentialModels}.

In this paper we analyze events with two photons and two leptons
(two electrons or two muons). The events are constrained to be
consistent with the $\Y(3S)\to\gamma\gamma\Y(2S)$ or 
$\Y(3S)\to\gamma\gamma\Y(1S)$ transitions, with the $\Y(2S)$ or $\Y(1S)$ decaying
to a lepton pair.
The analysis of similar events containing four photons and two leptons is the subject
of a separate paper submitted to this conference \cite{1dpaper}.

The various photon transitions that contribute to our data are outlined in 
Fig.~\ref{fig:levels}.
The two-photon sample is dominated by photon transitions from the $\Y(3S)$ to one 
of the triplet $\chi_b(2P_J)$ states, followed by  a subsequent photon transition
to the $\Y(2S)$ or $\Y(1S)$. 
The product branching ratios for these decays can be unfolded 
using the known measurements
for $\B(\chi_b(2P_J)\to\gamma\Y(2S))$ or $\B(\chi_b(2P_J)\to\gamma\Y(1S))$.
The ratios of these branching fractions for the same $J$ and
different final state $\Y$  test theoretical
predictions for the ratio of the corresponding E1 matrix elements, 
$|\!<\!2S|r|2P\!>\!|/|\!<\!1S|r|1P\!>\!|$.
Ratios of these branching fractions for different $J$ and the same 
final state $\Y$ give us insight into the ratios of the hadronic widths of the various $\chi_b(2P_J)$
states.
Measurement of the photon energies in the first transition
allows a mass determination of the $\chi_b(2P)$ states.

We also observe the rare photon cascade $\Y(3S)\to\gamma\chi_b(1P_J)$, $\chi_b(1P)\to\gamma\Y(1S)$.
From the measured product branching ratio,
we derive the $<\!1P|r|3S\!>$ E1 matrix element, which is of particular interest since 
different theoretical estimates of its value disagree.

We also use our sample to search for $\pi^0$ and $\eta$ transitions between the $\Y(3S)$ and
either the $\Y(2S)$ or $\Y(1S)$. The $\pi^0$ transitions are isospin violating decays.
Analogous transitions were previously observed in the $c\bar c$ system between the
$\psi(2S)$ and $J/\psi(1S)$ \cite{PDG}.

The two-photon cascade transitions via the $P$ states were previously observed 
by the CUSB \cite{CUSBexcl} and CLEO-II \cite{CLEOIIexcl} experiments.
Here, we present the results based on 1.1 fb$^{-1}$ of integrated luminosity accumulated 
at the $\Y(3S)$ resonance, corresponding to  $4.73\cdot 10^6$ $\Y(3S)$ decays.
This is roughly a ten-fold increase in
statistics compared to the CLEO-II data set, 
and roughly a four-fold increase compared to the integrated CUSB
data set.
Thanks to the good granularity and large solid angle 
of the CLEO CsI(Tl) calorimeter, the CLEO-III detection efficiency 
for these final states is a factor of 2-3 larger than in 
the CUSB detector.
Even though the CLEO-III calorimeter is essentially the same
as in the CLEO-II detector \cite{CLEOIIdetector}, 
the photon selection efficiency and
detector resolution were improved in the endcaps and in part of the
barrel calorimeter thanks to a new, lower-mass, tracking system  \cite{CLEOIIIDR} 
built for CLEO-III. 
Another change was the replacement of the time-of-flight system by a RICH
detector in the barrel part. 
Finally, the calorimeter endcaps were 
restacked and moved farther away from the interaction point to
accommodate a new, higher luminosity, interaction point optics.

\ssimplex{htbp}{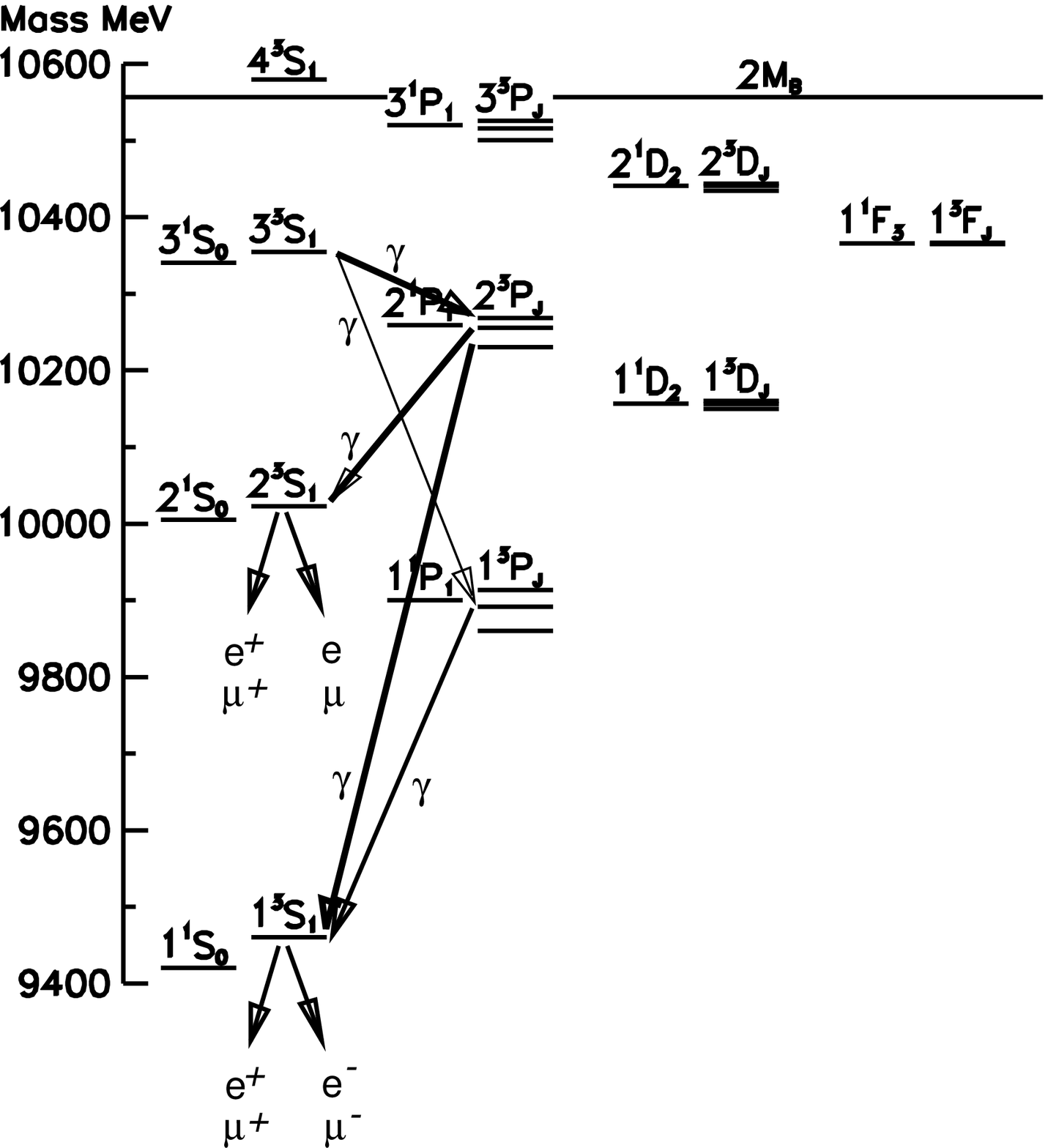}{levels}{
$\bb$ mass levels as predicted by one of the potential models.
The levels are denoted by their spectroscopic labels, 
$n^{2S+1}L_J$, where $n$ is the radial quantum number ($n=1, 2, \dots$),
$S$ is the total quark spin ($S=0$ spin singlets, $S=1$ spin triplets),
$L$ is the orbital angular momentum ($L=S, P, D, \dots$) and $J$ is the total
angular momentum of the state ($\vec{J}=\vec{S}+\vec{L}$).
Two-photon transition sequences via the $\chi_b(2P)$ or $\chi_b(1P)$ 
states are indicated.
}{10cm}

\section{Data selection}

We select events with exactly two photons and two oppositely
charged leptons. The leptons must have momenta of at least 3.75 GeV.
We distinguish between electrons and muons by their 
energy deposition in the calorimeter. Electrons must have a high
ratio of energy observed in the calorimeter to the momentum measured
in the tracking system ($E/p>0.7$).
Muons are identified as minimum ionizing particles, and are required to
leave $150-550$ MeV of energy in the calorimeter.
Stricter muon identification does not reduce the background
      in the final sample, since all significant background
      sources contain muons.
Each photon must have at least 60 MeV of energy. We also ignore
all photons below 180 MeV in the calorimeter region closest to the
beam, because of the spurious photons generated by beam-related backgrounds.
The total momentum of all photons and leptons in each event must be balanced 
to within 300 MeV.
The invariant mass of the two leptons must be consistent 
with either the $\Y(1S)$ or $\Y(2S)$ mass within $\pm300$ MeV.
Much better identification of the $\Y(1S)$ or $\Y(2S)$  resonance is
obtained by measuring  the mass of the system recoiling
against the two photons. The average resolution of the recoil mass is
15 MeV (9 MeV) for the $\Y(1S)$ ($\Y(2S)$). 
The resolution of the different states in the $\chi_b(2P)$ triplet
relies on the energy measurement of the lower energy photon in the event. 
This photon must be detected in the barrel 
part of the calorimeter, where the energy resolution is best.
The higher energy photon is allowed to be detected in the endcap
part for the $\gamma\gamma\mm$ events.
 
The final discrimination against the backgrounds is performed by using
the energy of the lower energy photon and the mass of the system recoiling against
the photons.
The energy of the less energetic photon, $E_{\gamma\,low}$, 
must peak for the signal events at one of the
values corresponding to $(M_{\Y(3S)}^2 - M_{\chib(2,1P_J)}^2)/(2\,M_{\Y(3S)})$
for various $J$.
The recoil mass, $M_{recoil}$, 
calculated using the photon four-vectors and the beam energy,
must peak at the $\Y(2S)$ or $\Y(1S)$ mass.
We actually use the difference between the center-of-mass energy
and the mass of the system recoiling against the photons, 
$\Delta M=E_{CM}-M_{recoil}$,
which peaks at $M_{\Y(3S)} - M_{\Y(2,1S)}$ 
for the signal events.
Scatter plots of $\Delta M$ versus $E_{\gamma\,low}$ 
are shown for $\mm$ and $\ee$ events separately
in Figs.~\ref{fig:mdrvseglowmm}-\ref{fig:mdrvseglowee}.
Clusters of events for $\Y(3S)\to\gamma\chib(2P_{2,1})$ followed by
a radiative decay to the $\Y(2S)$ or $\Y(1S)$ are clearly visible.
A less intense cluster of events, corresponding to 
$\Y(3S)\to\gamma\chib(1P_{2,1})$, 
$\chib(1P_{2,1})\to\gamma\Y(1S)$, is also visible.
The backgrounds vary smoothly over these two variables
and come predominantly from radiative Bhabhas and
$\mu-$pair events. The $\ee$ channel clearly has much higher
background due to the higher Bhabha cross-section.

\vduplex{tbhp}{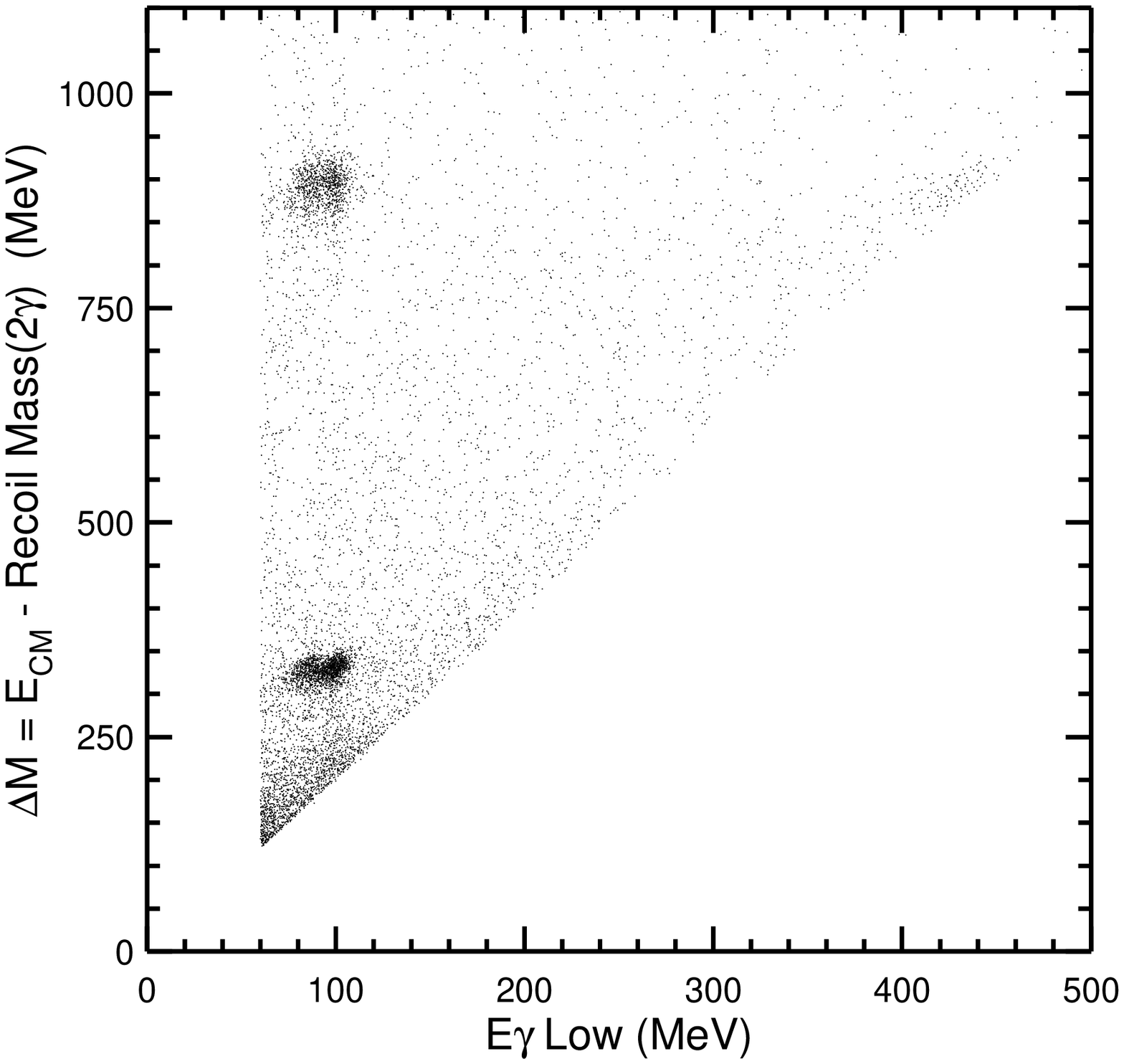}{mdrvseglowmm}{
The center-of-mass energy minus the mass of the system
recoiling against the two photons ($\Delta M$) vs.\ the energy of the less
energetic photon for $\gamma\gamma\mm$ events.
The two clusters of events correspond to
$\Y(3S)\to\gamma\chi_b(2P)$ ($E_{\gamma\,low}\sim90$ MeV)
followed by either
$\chi_b(2P)\to\gamma\Y(2S)$ ($\Delta M\sim330$ MeV) or
$\chi_b(2P)\to\gamma\Y(1S)$ ($\Delta M\sim895$ MeV).
A faint cluster of events corresponding to 
$\Y(3S)\to\gamma\chi_b(1P)$, $\chi_b(1P)\to\gamma\Y(1S)$
($E_{\gamma\,low}\sim410$ MeV, $\Delta M\sim895$ MeV), 
is also visible.}{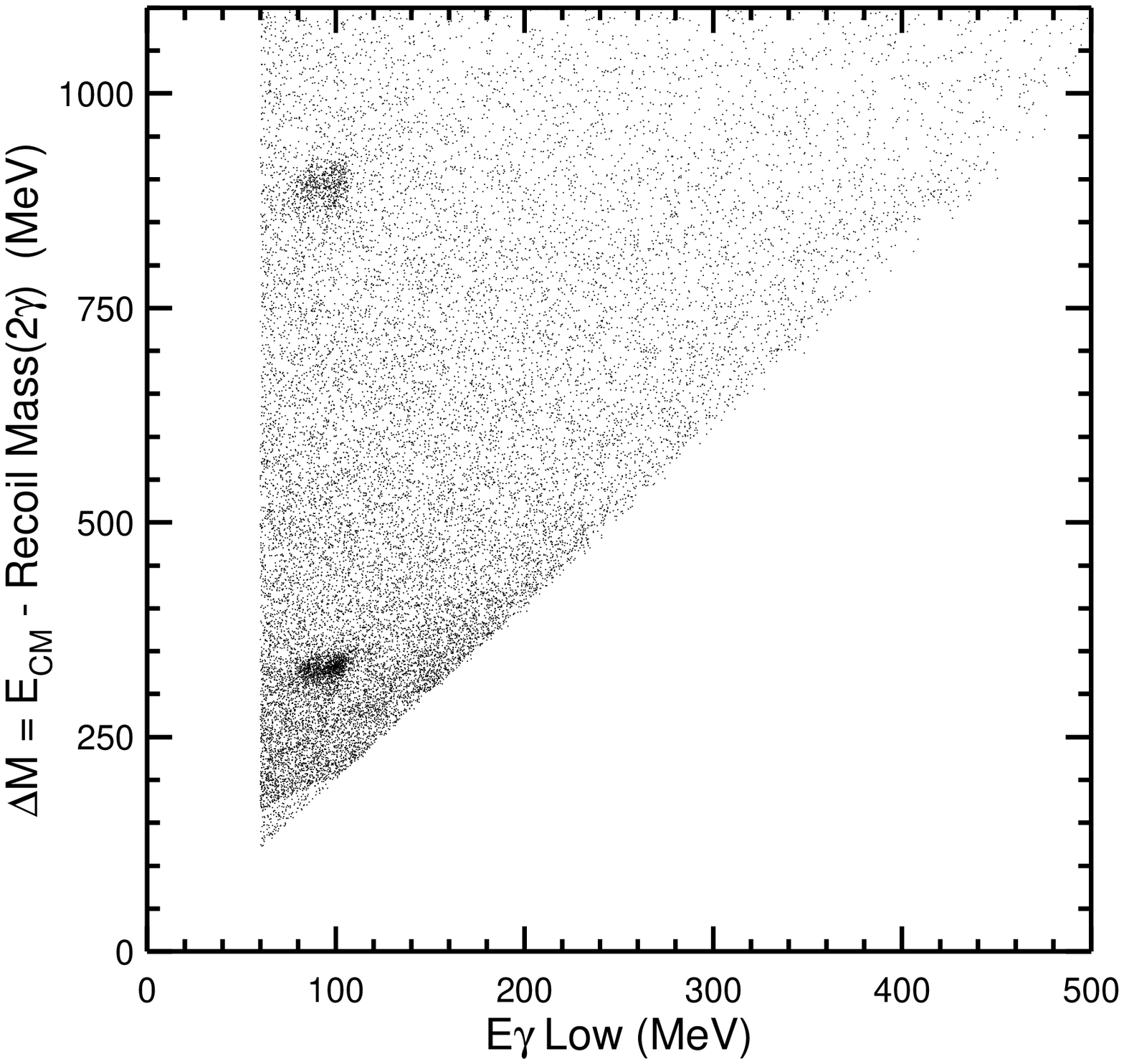}{mdrvseglowee}{
Similar plot for $\gamma\gamma\ee$ events.}

\section{Mass determination for the $\chib(2P_{2,1})$ states}
\label{sec:energies}

To determine the masses of the $\chi_b(2P)$ states, we combine the data for
the two-photon cascades to the $\Upsilon(2S)$ and the $\Upsilon(1S)$.
The recoil mass difference $\Delta M$ must be within $\pm3\sigma$ of
the expected value. 
The energy spectrum of the lower energy photon is plotted in
Fig.~\ref{fig:EGammaLow}.

\simplex{hbtp}{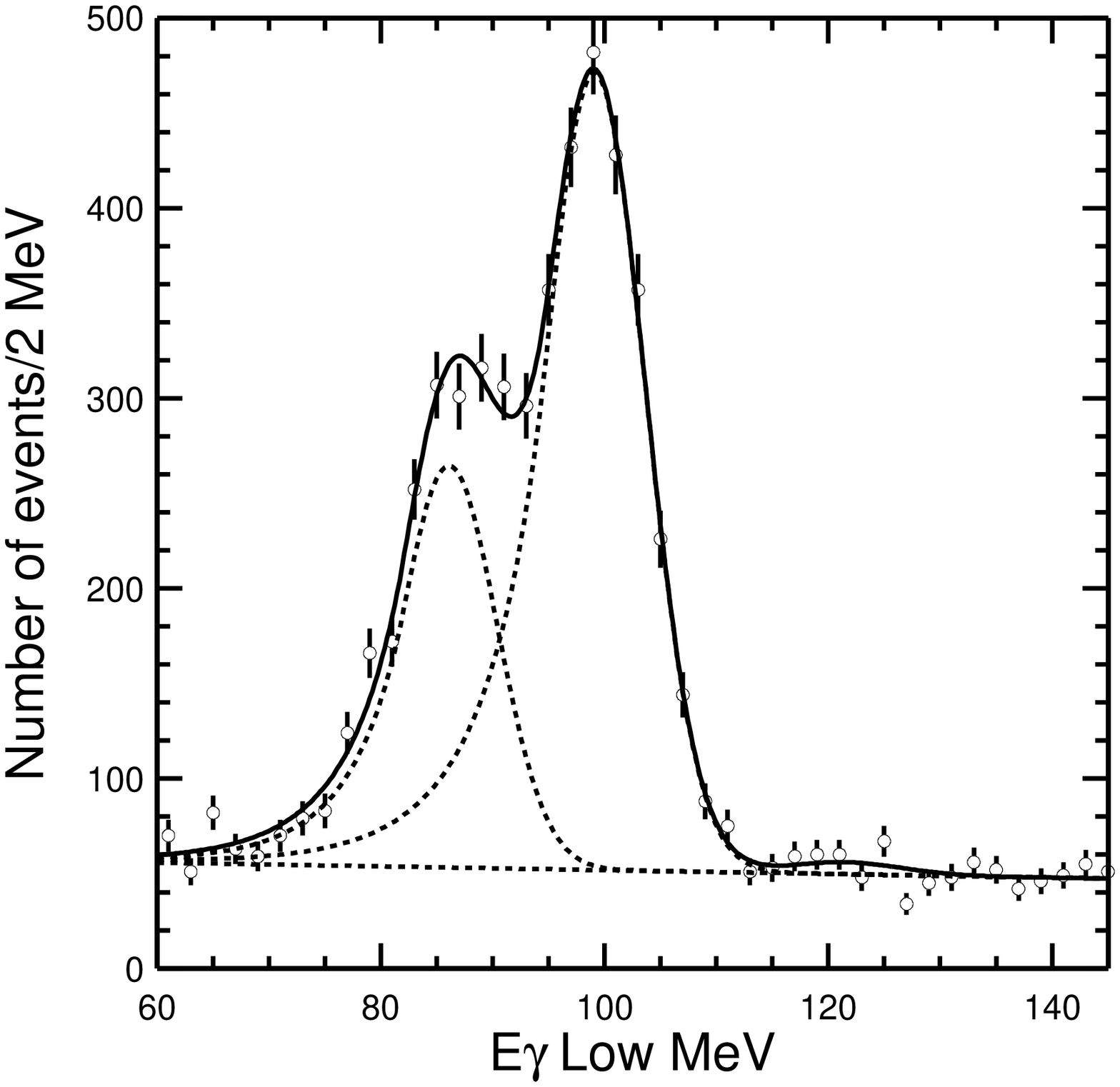}{EGammaLow}{
               Energy of the less energetic photon in
               $\Upsilon(3S)\to\gamma\gamma l^+l^-$ events.
               The two large peaks correspond to
               $\Upsilon(3S)\to\gamma\chi_b(2P_2)$ and
               $\Upsilon(3S)\to\gamma\chi_b(2P_1)$
               transitions. 
               There is also an insignificant peak at an energy
               corresponding to $\Upsilon(3S)\to\gamma\chi_b(2P_0)$. 
               The solid line represents the fit. The dashed lines
               show the fitted background (linear function) and the
               two individual photon lines on top of the background.
               }

From the three photon lines expected from 
$\Y(3S)\to\gamma\chi_b(2P_{2,1,0})$ transitions, only the
$J=2$ and $J=1$ lines are clearly visible.
The third $J=0$ line is suppressed by the larger hadronic width of
this state, which makes its radiative branching ratios small.
In the fit to the data (see Fig.~\ref{fig:EGammaLow}),
we parameterize each line 
by a Gaussian with an asymmetric low-energy tail. 
The number of events and the peak energy for each line are free parameters
in the fit, except for the position of the $J=0$ line, which was
fixed to the value obtained from the fit to the inclusive photon spectrum in
multi-hadronic events. 
The ratios of the widths of the three peaks were fixed to the Monte Carlo
determined dependence of the energy resolution on the photon energy.
The overall energy resolution scale factor was left free in the fit.
We also fixed the energy dependence of the turn-over point from the
Gaussian to the power-law tail from the Monte Carlo simulations.
Again, an overall scaling factor for the turn-over point was adjusted
to the data by the fit.
The backgrounds were assumed to be linear in energy, as suggested
by the $\Delta M$ sidebands, the off-resonance data 
(0.13 fb$^{-1}$ collected below the $\Y(3S)$ resonance)
and other data taken at the $\Y(4S)$.
The fit yields $1286\pm63$ events for the $J=2$ line positioned at
$86.1\pm0.3$ MeV, $2726\pm180$ events for the $J=1$ line positioned
at $99.1\pm0.2$ MeV and $45\pm34$ events for the $J=0$ line.
The fitted energy resolution scale factor is $1.05\pm0.03$, thus 
the data are  consistent with the Monte Carlo simulations.
The fitted energy resolution corresponds to $\sigma_{E_\gamma}=(4.6\pm0.2)$ MeV
for $E_\gamma=100$ MeV. 

Splitting the data into di-electron and di-muon subsamples, as well as
$\Y(2S)$ and $\Y(1S)$ subsamples, gives consistent results within the
statistical errors.
Varying the order of the background polynomial and the fit range
results in very small changes to the fitted photon  energies.
The dominant systematic error is due to
the uncertainties in fixing the absolute photon energy scale at various
photon energies. 
We used $\pi^0\to\gamma\gamma$ decays and the known $\pi^0$ mass
to correct for small non-linearities in the shower energy
determination.  
Then, we used the recoil mass distributions in the 
$\Y(3S)\to\gamma\gamma\Y(2S)$ and 
$\Y(3S)\to\gamma\gamma\Y(1S)$ data presented here, as well as 
the $\Y(3S)\to\pi^0\pi^0\Y(2S)$ and $\Y(3S)\to\pi^0\pi^0\Y(1S)$
transitions, together with the well known  
$\Y(1S)$, $\Y(2S)$, $\Y(3S)$ masses \cite{PDG},
to check the $\pi^0$ calibration.
From this study we have determined our systematic uncertainty in
the photon energies relevant for the $J=2,1$ lines
to be $0.35\%$.
 
Including the systematic error (strongly correlated between
the two lines), the photon energies are: 
$$
\begin{array}{c}
E_{J=2} = (86.09\pm0.30\pm0.29)\, \hbox{\rm MeV},\\
E_{J=1} = (99.08\pm0.17\pm0.34)\, \hbox{\rm MeV}.\\
\end{array}
$$
These results are consistent with, but more precise than, 
previous determinations \cite{PDG}.

Using the $\Y(3S)$ mass \cite{PDG}, 
we can turn these photon-line energies
into $\chi_b(2P_{2,1})$  masses:
$$
\begin{array}{c}
M_{\chi_b(2P_2)} = ( 10268.8\pm0.3\pm0.6)\, \hbox{\rm MeV},  \\
M_{\chi_b(2P_1)} = ( 10255.6\pm0.2\pm0.6)\, \hbox{\rm MeV}.  \\
\end{array}
$$

\section{Branching ratios for photon cascades via the $\chi_b(2P)$ states}

We can determine the product branching ratios for the cascade sequence
from the photon-line amplitudes fitted separately
to the $\Y(2S)$ and $\Y(1S)$ subsamples.
The efficiency for each photon line was determined from Monte Carlo simulations
that included the full angular correlations between the 
photons and leptons \cite{angular}. We also simulated final state radiation
in the annihilation to lepton pairs \cite{PHOTOS}.
In addition to one-dimensional fits to the $E_{\gamma\,low}$ distributions,
we also performed two-dimensional fits to $\Delta M$ vs.\ $E_{\gamma\,low}$
which improved the statistical errors.
The systematic errors were determined from variations of the cuts, and the fit procedure
and from the uncertainty in the number of $\Y(3S)$ decays in our sample.
After we determine the product branching ratios for $\B(\Y(3S)\to\gamma\gamma l^+l^-)$,
we also unfold them for  $\B(\Y(3S)\to\gamma\gamma\Y(2,1S))$ using
the world average values for $\B(\Y(2,1S)\to l^+l^-)$ \cite{bmm}, and finally,
for $\B(\chi_b(2P_J)\to\gamma\Y(2,1S))$ using 
the world average values for $\B(\Y(3S)\to\gamma\chi_b(2P_J))$ \cite{PDG}.
All the results are summarized in Table~\ref{tab:rates}, which also
contains a comparison with previous measurements.
The statistical significance of the $J=0$ signal for the $\Y(3S)\to\gamma\gamma\Y(2S)$ 
transitions is three standard deviations (see Table~\ref{tab:rates}).
Since the $J=0$ amplitude for the $\Y(3S)\to\gamma\gamma\Y(1S)$ transition is less
significant ($2.5\sigma$), we do not claim observation of 
this transition and set an upper limit on its rate at the 90\%~confidence level.
Our results are generally in agreement with previous
measurements, except for the cascade rates via
the $\chib(2P_1)$ state, for which we (and the previous CLEO-II
analysis) measure a substantially higher rate than obtained 
by CUSB. In all cases, our measurements have better precision than the
previous determinations.

\begin{table}[htbp]
\footnotesize
\begin{center}
\caption{Rate results for
$\Y(3S)\to\gamma\chib(2P_J)\to\gamma\gamma\Y(2,1S)\to\gamma\gamma\LL$.
$\B(\gamma\gamma\LL)$ means an average of 
$\B(\gamma\gamma\mm)$ and
$\B(\gamma\gamma\ee)$.
The first error is statistical and second 
(if given) is systematic.
CLEO-II and CUSB measurements for 
$\B(\gamma\gamma\LL)$ are also given for
comparison.
Upper limits are at the 90\%~C.L.
 \label{tab:rates}}
\def\1#1{\multicolumn{2}{c|}{#1}}
\renewcommand{\arraystretch}{1.5}
\begin{tabular}{|c|c|c|c|c|c|c|}
\hline
   & \1{$J=2$} & \1{$J=1$} & \1{$J=0$} \\
\cline{2-7}
   & $\mu\mu$ & $ee$ 
   & $\mu\mu$ & $ee$ 
   & $\mu\mu$ & $ee$ \\
\hline
\hline
    \multicolumn{7}{|c|}{$\Y(3S)\to\Y(2S)$} \\
\hline
Number of events 
   &   483$\pm$32 &  291$\pm$30 
   &  1081$\pm$39 &  697$\pm$36  
   &    $30.3^{+11.0}_{-10.1}$ & $21.0^{+17.4}_{-16.4}$ \\
\hline
Efficiency (\%)  
   &  36.5$\pm$0.4    & 23.7$\pm$0.4 
   &  38.2$\pm$0.4    & 26.3$\pm$0.4 
   &  39.3$\pm$0.4    & 23.7$\pm$0.4 \\
\hline
$\B(\gamma\gamma\LL)$ in $10^{-4}$ 
CLEO-III
   & \1{2.73$\pm$0.15$\pm$0.24} 
   & \1{5.84$\pm$0.17$\pm$0.41}
   & \1{0.17$\pm$0.06$\pm$0.02} \\
\hline
CLEO-II \cite{CLEOIIexcl}
   & \1{2.49$\pm$0.47$\pm$0.31} 
   & \1{5.11$\pm$0.60$\pm$0.63}
   & \1{$<$0.60} \\
\hline
CUSB  \cite{CUSBexcl}
   & \1{2.74$\pm$0.33$\pm$0.18} 
   & \1{3.30$\pm$0.33$\pm$0.19} 
   & \1{0.40$\pm$0.17$\pm$0.03}   \\
\hline
\hline
$\B(\Y(3S)\to\gamma\gamma\Y(2S))$ in $10^{-2}$ 
CLEO-III
   & \1{2.20$\pm$0.12$\pm$0.31} 
   & \1{4.69$\pm$0.14$\pm$0.62} 
   & \1{0.14$\pm$0.05$\pm$0.02} \\
\hline
\hline
$\B(\chi_b(2P)\to\gamma\Y(2S))$ in $10^{-2}$ 
CLEO-III  
   &  \1{19.3$\pm$1.1$\pm$3.1}
   &  \1{41.5$\pm$1.2$\pm$5.9}
   &  \1{2.59$\pm$0.92$\pm$0.51} \\
\hline
\hline
   \multicolumn{7}{|c|}{$\Y(3S)\to\Y(1S)$} \\
\hline
Number of events 
   &   353$\pm$25 &   186$\pm$23 
   &   587$\pm$28 &   370$\pm$26  
   & $18.0^{+7.7}_{-6.8}$ & $7.8^{+11.9}_{-10.8}$ \\
\hline
Efficiency (\%)  
   & 36.4$\pm$0.6    & 24.8$\pm$0.6 
   & 37.9$\pm$0.6    & 25.7$\pm$0.6 
   & 39.9$\pm$0.6    & 27.6$\pm$0.6 \\
\hline
$\B(\gamma\gamma\LL)$ in $10^{-4}$ 
CLEO-III
   & \1{1.93$\pm$0.12$\pm$0.17} 
   & \1{3.19$\pm$0.13$\pm$0.18} 
   & \1{$<0.16$} \\
\hline
CLEO-II \cite{CLEOIIexcl}
   & \1{2.51$\pm$0.47$\pm$0.32}
   & \1{3.24$\pm$0.56$\pm$0.41}
   & \1{$<$0.32}  \\
\hline
CUSB \cite{CUSBexcl}
   & \1{1.98$\pm$0.28$\pm$0.12}
   & \1{2.34$\pm$0.28$\pm$0.14}
   & \1{0.13$\pm$0.10$\pm$0.03} \\
\hline
\hline
$\B(\Y(3S)\to\gamma\gamma\Y(1S))$ in $10^{-2}$ 
CLEO-III
   & \1{0.79$\pm$0.05$\pm$0.07} 
   & \1{1.31$\pm$0.05$\pm$0.08}
   & \1{$<$0.08} \\
\hline
\hline
$\B(\chi_b(2P)\to\gamma\Y(1S))$ in $10^{-2}$ 
CLEO-III 
   &  \1{7.0$\pm$0.4$\pm$0.8}
   &  \1{11.6$\pm$0.4$\pm$0.9}
   &  \1{$<$1.44} \\
\hline
\end{tabular}
\end{center}
\end{table}

\section{Branching ratios for photon cascades via the $\chi_b(1P)$ states}

For $\Y(3S)\to\gamma\chib(1P)$, $\chib(1P)\to\gamma\Y(1S)$
transitions, the first and second photons have similar energies.
Furthermore, the photon energies of the $J=2$ and $J=1$  lines
cannot be resolved with
our energy resolution.
Therefore, the $E_{\gamma\, low}$ variable is no longer useful.
In this channel, we measure the sum of the branching fractions for
all the $J$ states. The $J=0$ contribution is expected to be
small since it is suppressed by the large hadronic 
width of this state.
To obtain a signal amplitude, we fit the $\Delta' M/\sigma(\Delta M)$ distribution, 
where $\Delta'M=\Delta M-(M_{\Y(3S)}-M_{\Y{1S}})$,
after requiring 
$E_{\gamma\, high}-E_{\gamma\, low}<100$ MeV. 
While for di-electron events both photons are required to be in the
barrel region of the calorimeter, for di-muon events 
one photon is allowed in the endcaps.
In the fits, the signal shape was determined from Monte Carlo events.
The fits are displayed in Figs.~\ref{fig:1pmm}-\ref{fig:1pee} and
the results are tabulated in Table~\ref{tab:1p}.
The systematic error were determined from variations of the cuts, and the fit procedure
and from the uncertainty in the number of $\Y(3S)$ decays in our sample.

\begin{table}[hbtp]
\footnotesize
\begin{center}
\caption{Results for
$\Y(3S)\to\gamma\chib(1P_J)\to\gamma\gamma\Y(1S)\to\gamma\gamma\LL$
summed over all the $J$ states. 
The first error is statistical and  second 
(if given) is systematic.
A CUSB measurement is also shown for comparison.
\label{tab:1p}}
\renewcommand{\arraystretch}{1.5}
\def\1#1{\multicolumn{2}{c|}{#1}}
\def\2#1{\multicolumn{1}{c|}{#1}}
\begin{tabular}{|c|r|r|}
\hline
   & \2{$\mu\mu$} & \2{$ee$} \\
\hline
Number of events 
   &   118$\pm$13 
   &   49$\pm$13 \\
\hline
Efficiency (\%)  
   &  \2{45.4$\pm$0.4} 
   &  \2{28.8$\pm$0.3} \\
\hline
$\B(\Y(3S)\to\gamma\gamma\LL)$ in $10^{-4}$ 
CLEO-III 
   & \1{0.520$\pm$0.054$\pm$0.052} \\
\hline
\hline
$\B(\Y(3S)\to\gamma\gamma\Y(1S))$ in $10^{-2}$ 
CLEO-III 
    & \1{0.214$\pm$0.022$\pm$0.021} \\
\hline
CUSB \cite{CUSBexcl}
   & \1{0.12$\pm$0.04$\pm$0.01} \\
\hline
\end{tabular}
\end{center}
\end{table}

\duplex{hbtp}{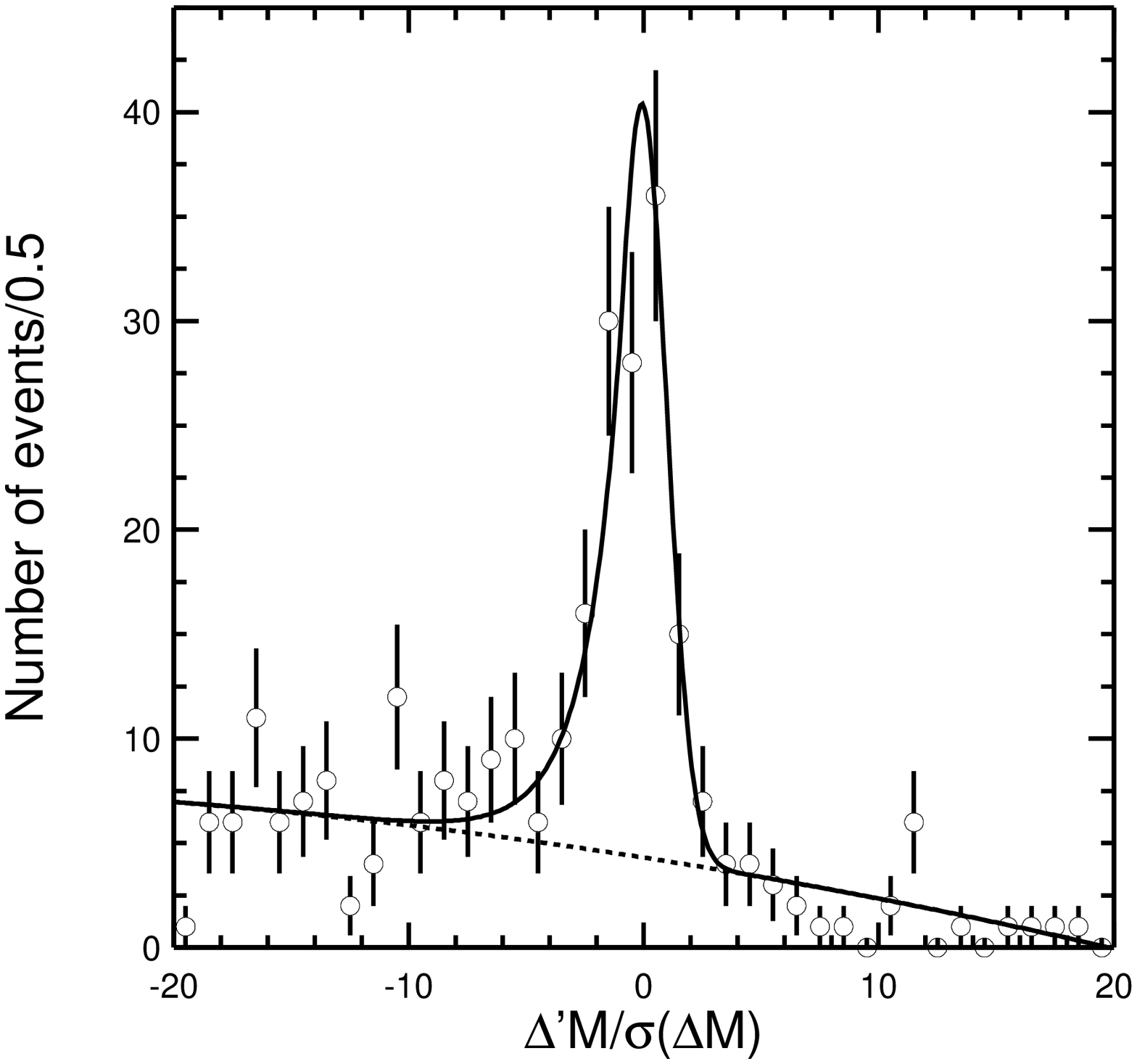}{1pmm}{
The $\Delta' M/\sigma(\Delta M)$ distribution 
for $\gamma\gamma\mm$ events.
               The solid line represents the fit. The dashed line
               shows the fitted background.
}{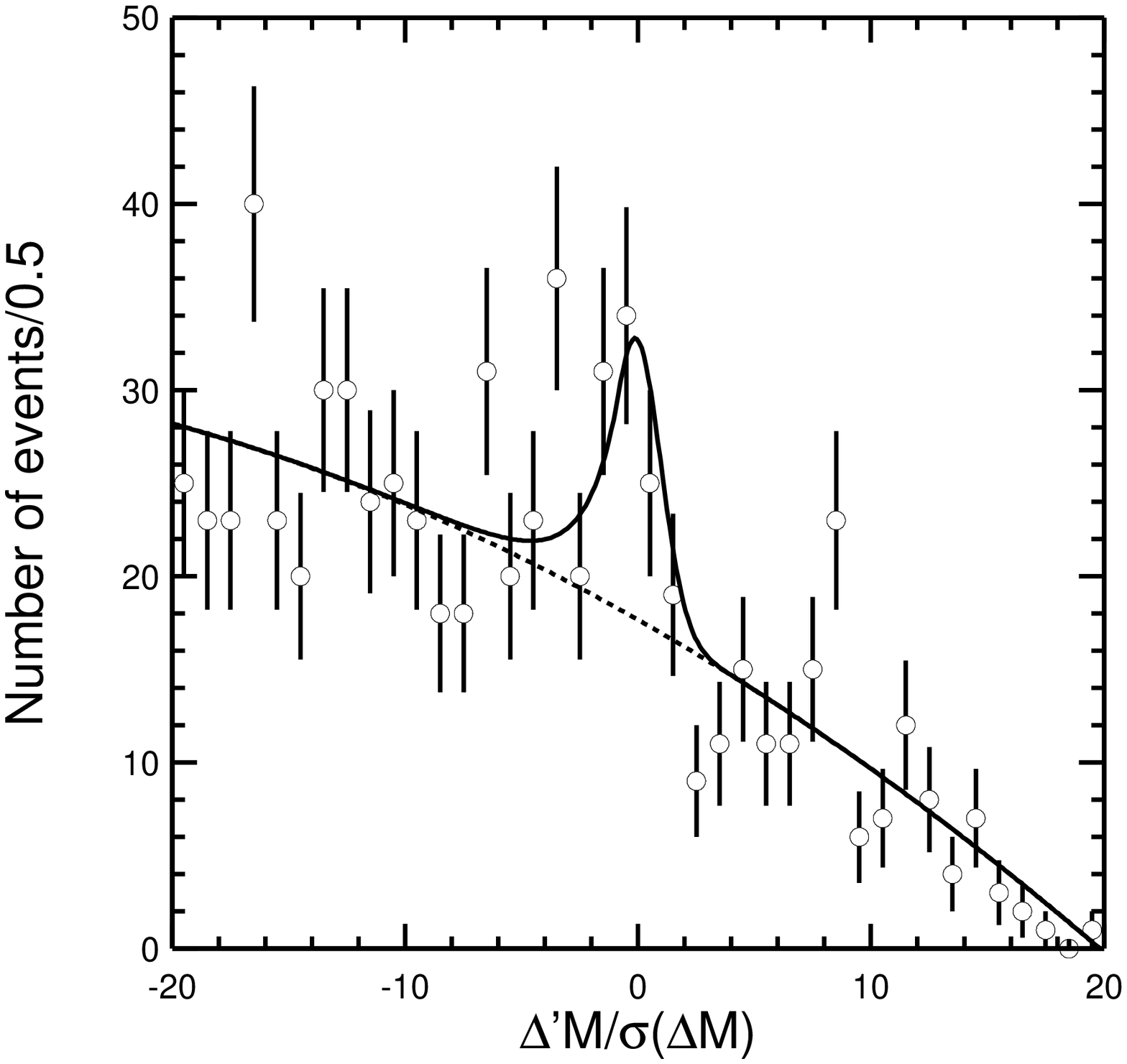}{1pee}{
The $\Delta' M/\sigma(\Delta M)$ distribution 
for $\gamma\gamma\ee$ events.}

\section{Search for $\pi^0$ and $\eta$ transitions}

\global\def\1{p02sfit}
\global\def\2{eta1sfit}
\global\def\3{p01s}
\global\def\4{p01sfit}
\quadruplex{htbp}{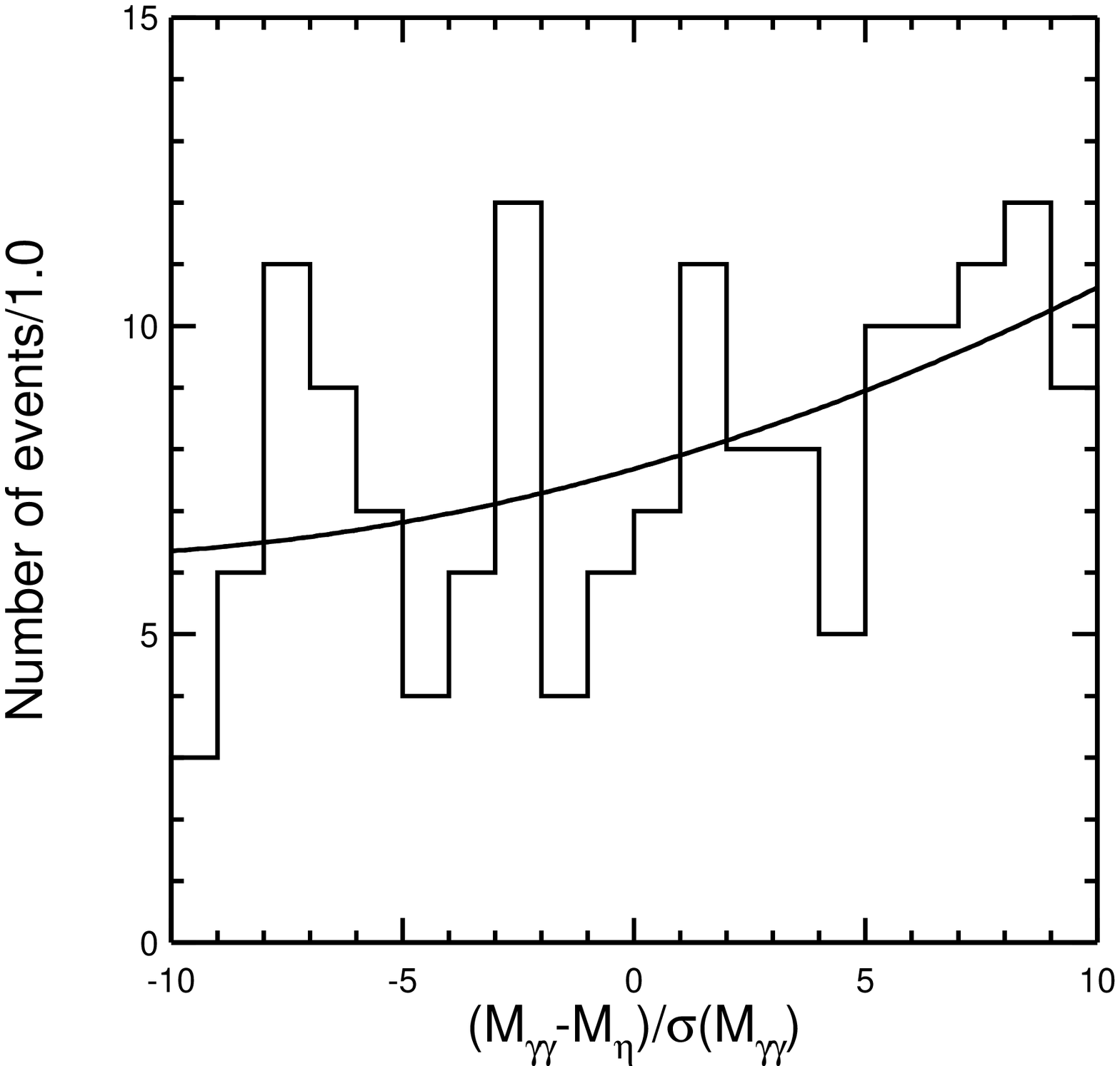}{
The deviation of the two-photon invariant mass from the
$\eta$ mass for $\Y(3S)\to\gamma\gamma\Y(1S)$ events.
The solid line represents the fit. The number of
signal events from the fit is zero.
}{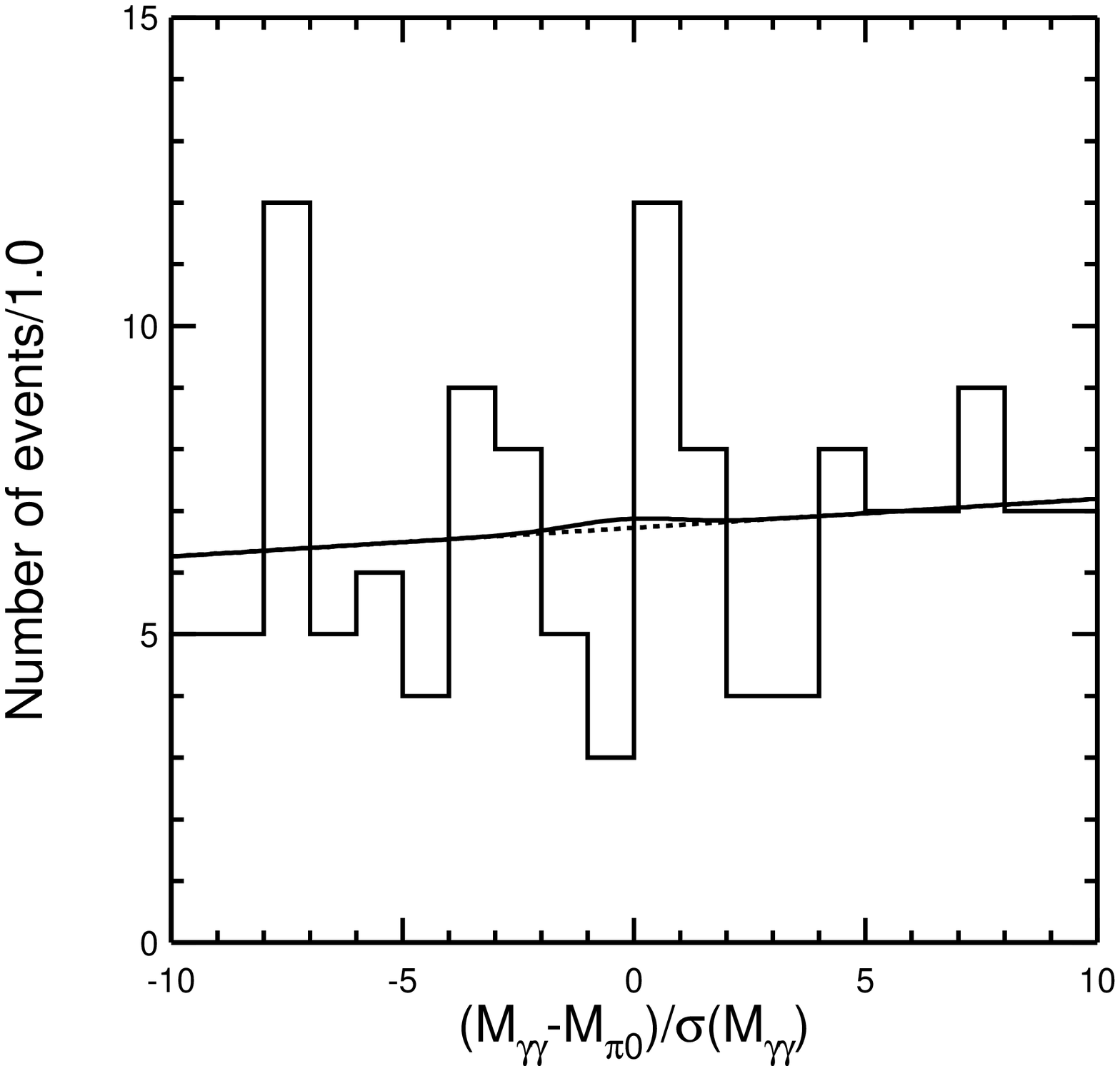}{
The deviation of the two-photon invariant mass from the
$\pi^0$ mass  for $\Y(3S)\to\gamma\gamma\Y(2S)$ events.
The solid line represents the fit. The dashed line
  represents the fitted background alone.
}{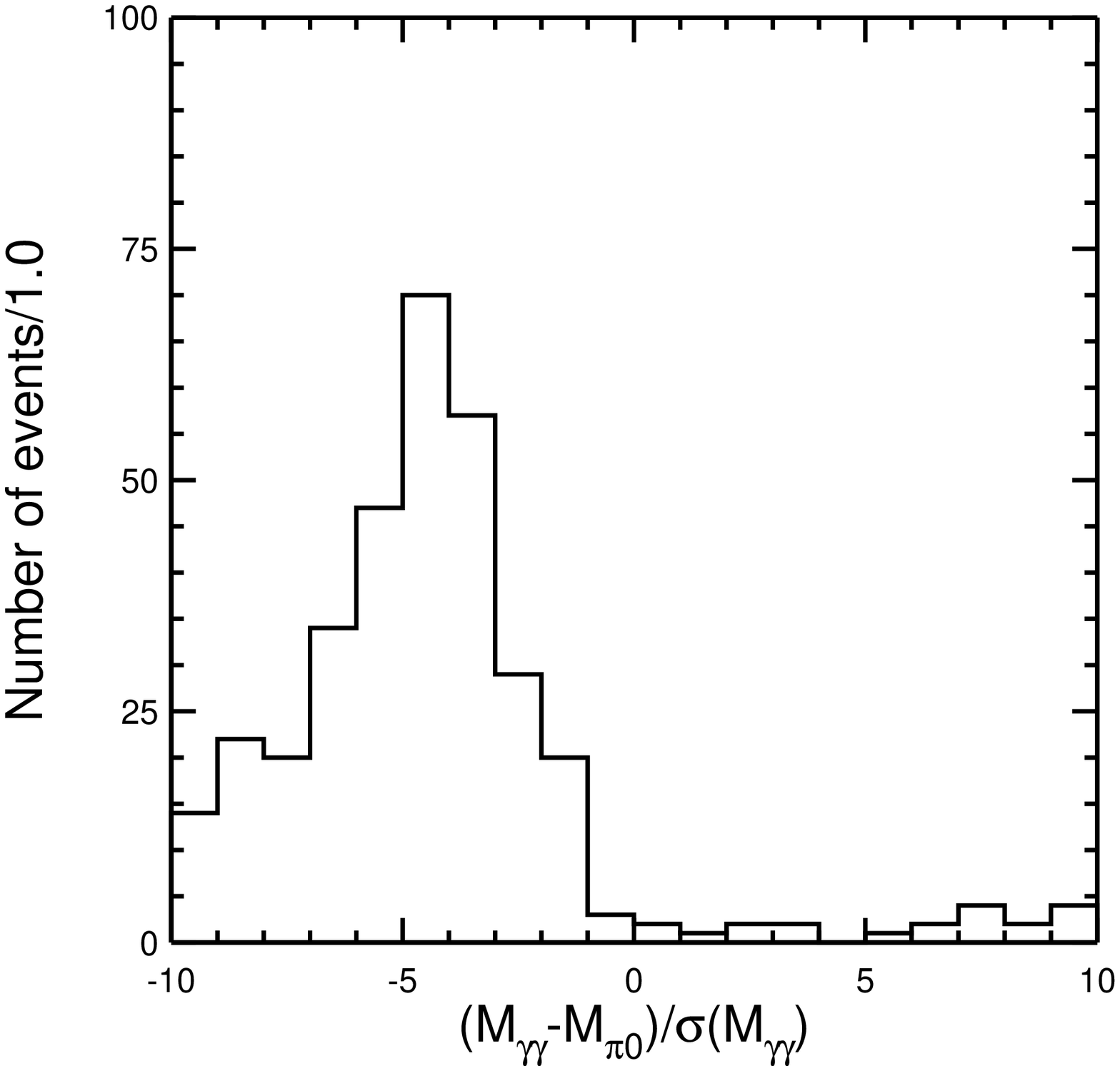}{
The deviation of the two-photon invariant mass from the
$\pi^0$ mass for $\Y(3S)\to\gamma\gamma\Y(1S)$ events.
}{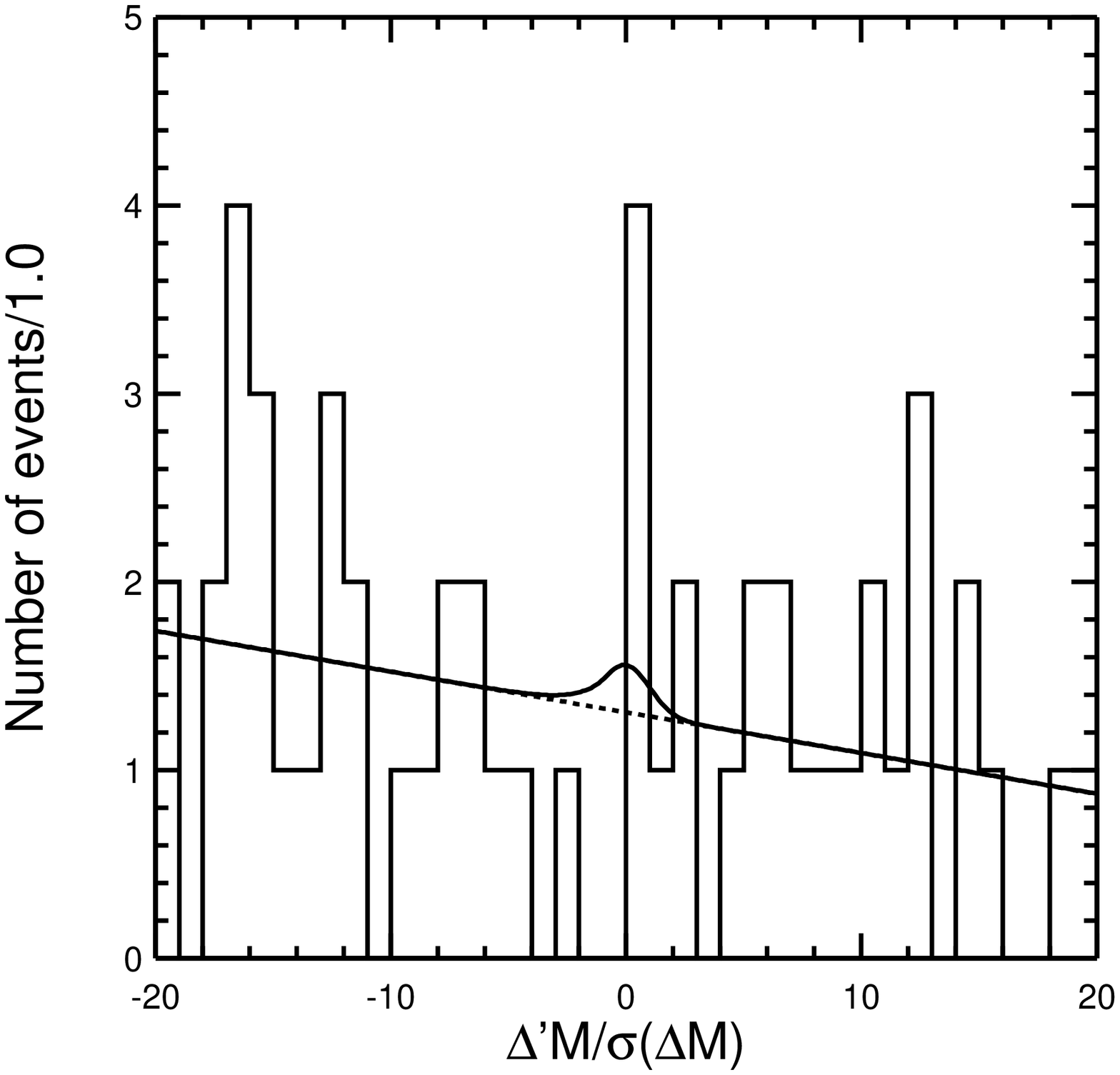}{
The recoil mass deviation from the $\Y(1S)$ mass for
$\Y(3S)\to\gamma\gamma\Y(1S)$ events.
The solid line represents the fit. The dashed line
  represents the fitted background alone.
}

To search for 
$\Y(3S)\to\pi^0\Y(2S)$, 
$\Y(3S)\to\pi^0\Y(1S)$ 
and $\Y(3S)\to\eta\Y(1S)$ transitions,
we use the same sample of events as for the studies
of the two-photon transitions described above.
To suppress the two-photon transitions, we
require $E_{\gamma\,low}>135$ MeV and
$E_{\gamma\,high}-E_{\gamma\,low}>100$ MeV for
the $\Y(1S)$ subsample, and 
$E_{\gamma\,low}>115$ MeV for
the $\Y(2S)$ subsample.
For $\Y(3S)\to\pi^0\Y(2S)\,$ 
$[\Y(3S)\to\eta\Y(1S)$]
we require the two-photon recoil mass to
be within 3 standard deviations of the 
$\Y(2S)\,$ $[\Y(1S)]$ mass,
and look for a $\pi^0\,$ $[\eta]$ mass peak in
the distribution of 
$(M_{\gamma\gamma}-M_{\pi^0})/\sigma(M_{\gamma\gamma})\,$
$[(M_{\gamma\gamma}-M_{\eta})/\sigma(M_{\gamma\gamma})]$,
where $\sigma(M_{\gamma\gamma})$ is the expected mass
resolution for a given $\gamma\gamma$ pair.
The signal is expected to be nearly the standard Gaussian
(mean close to zero and $\sigma\approx1$)
in these variables. No such peaks are observed in the data,
as shown in Figs.~\ref{fig:p02sfit}-\ref{fig:eta1sfit}.
From the fit of a signal contribution on top of a polynomial
background, we obtain upper limits on the number of 
observed signal events.
For the $\Y(3S)\to\pi^0\Y(1S)$ candidates, the backgrounds
peak sharply below the $\pi^0$ mass, as shown in Fig.~\ref{fig:p01s},
which makes it difficult to fit this distribution for
a signal peak. Therefore, instead we first require
$-1<(M_{\gamma\gamma}-M_{\pi^0})/\sigma(M_{\gamma\gamma})<+3$
and then plot the $\pi^0$ recoil mass distribution  
(see Fig.~\ref{fig:p01sfit}).
The backgrounds are smooth in this variable.
Again, no signal peak is observed and we set an upper limit
on the number of signal events.
The systematic error in the efficiency determination and 
the signal extraction is found to be 20\%.
We increase the upper limits on the branching ratios by 20\%\ to
account for this uncertainty.
The results are summarized in Table~\ref{tab:p0eta}.

\begin{table}[hbtp]
\footnotesize
\begin{center}
\caption{Results for
$\Y(3S)\to\pi^0\Y(1S)$,
$\Y(3S)\to\eta\Y(1S)$,
$\Y(3S)\to\pi^0\Y(2S)$
transitions. 
The efficiency quoted for the $\eta$
transition excludes 
$\B(\eta\to\gamma\gamma)$ \cite{PDG}.
\label{tab:p0eta}}
\renewcommand{\arraystretch}{1.5}
\begin{tabular}{|c|c|c|c|}
\hline
 &
$\Y(3S)\to\pi^0\Y(1S)$ &
$\Y(3S)\to\eta\Y(1S)$  &
$\Y(3S)\to\pi^0\Y(2S)$ \\
\hline
\#~of signal events & $0.7^{+2.8}_{-0.7}$ &
                      $0.0^{+3.5}_{-0.0}$ &
                      $0.4^{+6.5}_{-0.4}$ \\
\hline
90\%~C.L.~U.L.~on signal events &
                      $<5.3$ &
                      $<6.0$ &
                      $<11.0$ \\
\hline
efficiency (\%) & $16.5\pm0.4$ & 
                  $8.9\pm0.3$ &
                  $9.7\pm0.3$ \\
\hline
90\%~C.L.~U.L.~on $\B(\pi^0/\eta\LL)$ &
                  $<0.41\cdot 10^{-5}$ &
                  $<2.2\cdot 10^{-5}$ &
                  $<1.5\cdot 10^{-5}$ \\
\hline
90\%~C.L.~U.L.~on $\B(\pi^0/\eta)$ &
                  $<0.17\cdot 10^{-3}$ &
                  $<0.90\cdot 10^{-3}$ &
                  $<1.2\cdot 10^{-3}$ \\
\hline\hline
previous U.L.~on $\B(\eta)$ \cite{3seta} &
         --- & $<2.2\cdot10^{-3}$ & --- \\
\hline
\end{tabular}
\end{center}
\end{table}

\section{Comparison of the two-photon transition results to theory}

\subsection{E1 matrix elements}

The $\B(\Y(3S)\to\gamma\gamma\Y(2,1S))$ via $\chib(2P)$ and $\chib(1P)$ states
measured here can be compared to potential model predictions for the E1 matrix
elements in two ways. First,
the ratio of the magnitudes of the matrix elements for the decay of the same
$\chib(2P_J)$ state to different $S$ states can be obtained from:

\def\mratio#1{\frac{ \eonem{2P_#1}{1S} }{ \eonem{2P_#1}{2S} }}
$$
\mratio{J} =
\sqrt{  \frac{ \B(3S\to\gamma 2P_J, 2P_J\to\gamma 1S) }{ 
               \B(3S\to\gamma 2P_J, 2P_J\to\gamma 2S) }
        \left( \frac{E_\gamma(2P_J\to 2S)}{E_\gamma(2P_J\to 1S)} \right)^3 
      },
$$
where $E_\gamma(2P_J\to 1,2S)$ can be obtained from the well known masses
of the initial  and final states \cite{PDG}.
Using the branching ratios given in Table~\ref{tab:rates}, we obtain:
$$
\renewcommand{\arraystretch}{2.0}
\begin{array}{rl}
\mratio{2}&=0.105\pm0.004\pm0.006, \\
\mratio{1}&=0.087\pm0.002\pm0.005, \\
\mratio{2}/\mratio{1}&=1.21\pm0.06, \\
\mratio{{2,1}}&=0.096\pm0.002\pm0.005, \\
\end{array}
$$
where the first error is due to the statistical uncertainties in the 
radiative branching ratios and the second error is due to
the uncertainties in $\B(\Y(1,2S)\to\LL)$ (thus, we assume that all other
systematics cancel). 
In the non-relativistic limit, the E1 matrix elements do not
depend on $J$. Since our results for $J=2$ and $J=1$
differ by 3.5 standard deviations, we conclude that there is
evidence for relativistic effects.
To compare to the potential model prediction, we
calculated an average over $J=2$ and $J=1$ (the fourth result above).

We can also extract the $ \eonem{1P}{3S} $ matrix element from
the photon transitions via the $\chib(1P)$ states:
$$
 \eonem{1P}{3S}  =
\sqrt{
      \frac{ \B(3S\to\gamma 1P, 1P\to\gamma 1S)\,\Gamma_{tot}(3S) }{
           { D\, \sum_J (2J+1) E_\gamma(1P_J\to 1S)^3 \B(1P_J\to\gamma 1S) }}
} \quad,       
$$
where $D=4/27\,\alpha\,{e_b}^3=1.2\cdot 10^{-4}$ ($\alpha$ is the fine 
structure constant, and $e_b$ is the $b-$quark electric charge).
This formula assumes that the matrix element is spin independent.
Taking $\B(3S\to\gamma 1P, 1P\to\gamma 1S)$ from Table~\ref{tab:1p}
and the world average values for the other quantities \cite{PDG}, 
we obtain:
$$
 \eonem{1P}{3S}  = (0.050\pm0.006)\, GeV^{-1}\,.
$$
The error here includes the statistical and systematic uncertainties on all
quantities added in quadrature.

These results are compared to various potential model predictions
in Table~\ref{tab:radmatrix}.
We also include a comparison between various potential model
predictions and experimental results for 
$ \eonem{2P}{3S} $ and 
$ \eonem{1P}{2S} $
extracted from 
the world average
results for $B(\Y(3S)\to\gamma\chib(2P_J))$ and
$B(\Y(2S)\to\gamma\chib(1P_J)$ 
(taken from Ref.~\cite{BessonSkwarnicki}).

\begin{table}
\caption{\label{tab:radmatrix}
Comparison of E1 matrix elements and their ratios
predicted by different potential models
with measurements from $\bb$ data. ``NR'' denotes non-relativistic 
calculations and ``rel'' refers to models with relativistic corrections.}
\def\1#1{\multicolumn{1}{c}{#1}}
\def\2#1#2{\multicolumn{#1}{c|}{#2}}
\def\3#1#2{\multicolumn{#1}{c|}{#2}}
\def\etal{et al.}
\begin{center}
\begin{tabular}{|l|cc|cc|cc|cc|}
\hline
      & \2{2}{$|<2P|r|3S>|$} & \2{2}{$|<1P|r|2S>|$} &
     \2{2}{$|<1P|r|3S>|$} & \3{2}{$|<1S|r|2P>|$} \\
      & \2{2}{\quad}       & \2{2}{\quad}       & 
     \2{2}{\quad}  & \3{2}{$\overline{|<2S|r|2P>}|$}  \\
\cline{1-9} 
      & \2{2}{$GeV^{-1}$} & \2{2}{$GeV^{-1}$} & \2{2}{$GeV^{-1}$} &
              \3{2}{\quad} \\
\hline\hline
DATA &  \2{2}{2.7$\pm$0.2} & \2{2}{1.9$\pm$0.2} & \2{2}{0.050$\pm$0.006} 
     & \3{2}{0.096$\pm$0.005}   \\
\cline{2-9}
     & \multicolumn{4}{c|}{World Average} & \multicolumn{4}{c|}{This measurement} \\
\hline\hline
Model & NR & rel & NR & rel & NR & rel & NR & rel \\
\hline
Kwong, Rosner \cite{KwongRosner}     
& 2.7 &     & 1.6 &     & 0.023 &       & 0.13 &      \\
Fulcher \cite{Fulcher90}         
& 2.6 &     & 1.6 &     & 0.023 &       & 0.13 &      \\
B\"uchmuller \etal \cite{BuchmullerTye80} 
& 2.7 &     & 1.6 &     & 0.010 &       & 0.12 &      \\
Moxhay, Rosner \cite{Moxhay83}    
& 2.7 & 2.7 & 1.6 & 1.6 & 0.024 & 0.044 & 0.13 & 0.15 \\
Gupta \etal \cite{Gupta86}      
& 2.6 &     & 1.6 &     & 0.040 &       & 0.11 &      \\
Gupta \etal \cite{Gupta82}      
& 2.6 &     & 1.6 &     & 0.010 &       & 0.12 &      \\
Fulcher \cite{Fulcher88r}         
& 2.6 &     & 1.6 &     & 0.018 &       & 0.11 &      \\
Danghighian \etal \cite{Daghighian}
& 2.8 & 2.5 & 1.7 & 1.3 & 0.024 & 0.037 & 0.13 & 0.10 \\
McClary, Byers \cite{McClaryByers83}    
& 2.6 & 2.5 & 1.7 & 1.6 &       &       & 0.15 & 0.13 \\
Eichten \etal  \cite{Eichten80}  
& 2.6 &     & 1.7 &     & 0.110 &       & 0.15 &      \\
Grotch \etal \cite{Grotch84}     
& 2.7 & 2.5 & 1.7 & 1.5 & 0.011 & 0.061 & 0.13 & 0.19 \\
\hline
\end{tabular}
\end{center}
\end{table}

While  most of the potential models have no trouble 
reproducing the large matrix elements,
$ \eonem{2P}{3S} $,
$ \eonem{1P}{2S} $,
which show also little model dependence,
only a few models predict 
$ \eonem{1P}{3S} $
in agreement with our measurement.
Clearly, the latter transition is more
sensitive to the underlying description
of $b\bar b$ states.
Predictions for the ratio
$\mratio{{\,\!}}$ are not as model dependent, but
somewhat higher than our experimental value.

\subsection{Ratio of hadronic widths of the $\chib(2P_J)$ states}

Assuming spin independence of the matrix elements,
the ratios of hadronic widths ($\Gamma_{had}=\Gamma_{tot}-\Gamma_{E1}$)
of the $\chib(2P_J)$ states can be
derived from:
$$
\frac{\Gamma_{had}(2P_{Ja})}{\Gamma_{had}(2P_{Jb})}=
\left( \frac{E_\gamma(2P_{Ja}\to 2S)}{E_\gamma(2P_{Jb}\to 2S)} \right)^3
\frac{1/\B(2P_{Ja}\to\gamma 2S)\,-1}{1/\B(2P_{Jb}\to\gamma 2S)\,-1}\quad.
$$

Taking $\B(2P_{Ja}\to\gamma 2S)$ from Table~\ref{tab:rates} and
applying this formula to $Ja=0$ and $Jb=2$, we obtain:
$$
\frac{\Gamma_{had}(2P_{0})}{\Gamma_{had}(2P_{2})}=5.6\pm2.6.
$$
To leading order in perturbative QCD, both states annihilate
to two hard gluons and the ratio is expected to be simply the ratio of
spin counting factors: 15/4=3.75 \cite{hadronicWidths}.
However, QCD correction are known to be large. Thus, there is rough agreement
between the data and the expectations. A more precise determination of
$\B(2P_{0}\to\gamma 2S)$ would be useful.

The $J=1$ state cannot annihilate to two massless gluons. Thus, to
leading order in perturbative QCD, the hadronic width of the $2P_1$ state
is suppressed by one power of the strong coupling constant.
The data confirm this suppression:
$$
\frac{\Gamma_{had}(2P_{1})}{\Gamma_{had}(2P_{2})}=0.29\pm0.06.
$$

\section{Summary}

We have presented improved measurements of 
the photon energies in 
$\Y(3S)\to\gamma\chi_b(2P_2)$ and 
$\Y(3S)\to\gamma\chi_b(2P_1)$.
We have also produced improved measurements of
the product branching ratios for
$\Y(3S)\to\gamma\gamma\Y(2S)$ and 
$\Y(3S)\to\gamma\gamma\Y(1S)$ 
proceeding through the $\chi_b(2P_J)$ states (measured
for each $J$ separately), and proceeding through
the $\chi_b(1P_J)$ states (measured as a sum
for $J=2,1$).

Finally, we have determined upper limits on the branching ratios for 
$\Y(3S)\to\pi^0\Y(2S)$, $\Y(3S)\to\pi^0\Y(1S)$, and
$\Y(3S)\to\eta\Y(1S)$, the first two of which were obtained for the first time.

We gratefully acknowledge the effort of the CESR staff in providing us with
excellent luminosity and running conditions.
M. Selen thanks the PFF program of the NSF and the Research Corporation, 
and A.H. Mahmood thanks the Texas Advanced Research Program.
This work was supported by the National Science Foundation, and the
U.S. Department of Energy.


\begin{thebibliography}{99}
\bibitem{bmm}
We used $\B(\Y(1S)\to\LL)=0.0243\pm0.0006$, 
$\B(\Y(2S)\to\LL)=0.01245\pm0.0014$, which are averages
of the $\mm$ and $\ee$ world-average branching ratios \cite{PDG}.
\end{thebibliography}
\end{document}